\documentclass{aa}  
\usepackage{graphicx}
\usepackage{txfonts}
\usepackage{url}
\usepackage{natbib}
\usepackage[utf8]{inputenc}
\usepackage[breaklinks=true]{hyperref} 
\hypersetup{colorlinks=true,urlcolor=blue,citecolor=blue,pdfborder= 0 0 0}
\bibpunct{(}{)}{;}{a}{}{,}    
\begin{document}

   \title{Merging groups and clusters of galaxies from the SDSS data}

   \subtitle{The catalogue of groups and potentially merging systems}

   \author{E.~Tempel\inst{1,2} \and T.~Tuvikene\inst{2} \and
           R.~Kipper\inst{2} \and N.~I.~Libeskind\inst{1}
          }

   \institute{Leibniz-Institut f\"ur Astrophysik Potsdam (AIP), An der Sternwarte 16, 14482 Potsdam, Germany\\
   \email{etempel@aip.de}
   \and Tartu Observatory, Observatooriumi 1, 61602 T\~oravere, Estonia}

   \date{}

 
  \abstract
   {Galaxy groups and clusters are the main tools used to test cosmological models and to study the environmental effect of galaxy formation.}
   {This work  provides a catalogue of galaxy groups and clusters, as well as potentially merging systems based on the SDSS main galaxy survey.}
   {We identified galaxy groups and clusters using the modified friends-of-friends (FoF) group finder  designed specifically for flux-limited galaxy surveys. The FoF group membership is refined by multimodality analysis to find subgroups and by using the group virial radius and escape velocity to expose unbound galaxies. We look for merging systems by comparing distances between group centres with group radii.}
   {The analysis results in a catalogue of 88\,662 galaxy groups with at least two members. Among them are 6873 systems with at least six members which we consider to be more reliable groups. We find 498 group mergers with up to six groups. We performed a brief comparison with some known clusters in the nearby Universe, including the  Coma cluster and Abell~1750. The Coma cluster in our catalogue is a merging system with six distinguishable subcomponents. In the case of Abell~1750 we find a clear sign of filamentary infall toward this cluster. Our analysis of mass-to-light ratio ($M/L$) of galaxy groups reveals that $M/L$ slightly increases  with group richness.}
   {}


   \maketitle
%

\section{Introduction}

It has been known since the very first large-scale structure surveys that galaxies cluster, due to their mutual gravitational attraction. This clustering of galaxies is known to have an effect on galactic properties since a galaxy's ability to access a reservoir of gas for star formation is regulated by  its environment (among
other factors). For example, large galaxies are seldom found in isolation. Empirical relations between morphology and environment have been known at least since \citet{1980ApJ...236..351D} identified the trend for early-type elliptical galaxies to inhabit, on average, denser environments than late type spirals. It is thus believed that a formal study of galaxy environment could uncover the main mechanisms of galaxy formation and evolution.

The definition of `environment' thus requires firm footing. Although many sophisticated methods for quantifying the environment exist in the literature (for example studies focusing on the topology of the  density, tidal, or velocity fields), there is still much to be learned from a simple approach. Perhaps the most intuitive way of quantifying a galaxy's environment is to  ascertain whether or not a given galaxy can be associated with a group of galaxies. 

There have been many different studies aimed at producing group catalogues from survey data, specifically from the SDSS. The first and possibly most cited of these catalogues is the NYU-Value Added Catalogue \citep{2005AJ....129.2562B} and its application to subsequent data releases. The method used in the present paper is based on previous group catalogues extracted from SDSS~DR10 \citep{2014A&A...566A...1T}. Further studies include a group finder with membership refinement, where members are iteratively removed to ensure a certain degree of gravitational binding among group members \citep{2016A&A...588A..14T}. A comparison of various group finders are presented in \citet{2014MNRAS.441.1513O, 2015MNRAS.449.1897O}.

These group finders serve  not only to quantify how light is distributed throughout the Universe, but also to estimate how this light is traced by the underlying distribution of matter since together these data may give an estimate for cosmological parameters such as the matter density of the Universe \citep[see][]{2014MNRAS.439.2505B}. Other studies of the mass-to-light ratio ($M/L$) obtained by studying groups of galaxies include \citet{2007A&A...464..451P} and 
\citet{2015MNRAS.449.2345P}, although it can be hard to make a direct comparison with between these studies because of differences in how the analyses are performed \citep[see also][for other estimates of the $M/L$ of galaxy groups]{2002ApJ...569..720G, 2009ApJS..183..197W}.

On the other hand, the accepted paradigm of the formation of  cosmic structures  suggests that they form hierarchically \citep{1978MNRAS.183..341W}:  small structures merge to form bigger ones and so on in a `bottom-up' way. Thus the violent collision and merging of structures is asserted as a dominant physical process at all scales and throughout cosmic time. Although cosmological numerical simulations suggest that there is a spectrum of mergers happening at any given time, simple idealized merging situations \citep[such as the timing argument;][]{1959ApJ...130..705K,2013MNRAS.436L..45P} have been invoked for such ambitious aims as measuring the age or dark matter content of the Universe. 

In numerical simulations it is relatively easy to identify merging structures since not only is the full cosmological evolution of each group readily accessible, but so too are indicators of the merging process, such as centre-of-mass displacement or the presence of a large number of unbound particles. In observations this is not the case since one is often dealing with flux limited surveys and sparse sampling. Recently, studies such as \citet{2016arXiv160904121Z} have gone to great lengths to produce simulated merger catalogues for the purpose of identifying the underlying physical processes present in puzzling observations. There are very few methods that study the automated substructure identification in observed redshift catalogues:  \citet{Yu:2015} investigated the power of a caustic technique to identify substructure of galaxy clusters, \citet{deLosRios:16} developed the mixture of Gaussian technique to identify merging systems, and  \citet{Wen:12} used
 group shape parameters to identify substructure.

A full catalogue of merging groups, based on a survey as large and deep as the SDSS has  not yet been published, although several single merging systems in these surveys have been discovered. For example, \citet{2016A&A...594A..32T} conducted an X-ray survey of SDSS stripe 82 and identified two merging clusters; \citet{2013A&A...557A..62P,2014A&A...570A..40P} identified individual merging systems in the southern sky; \citet{Flin:06} detected substructure on a sample of Abell clusters; and \citet{Ramella:07} analysed the substructures in WIde-field Nearby Galaxy-cluster Survey (WINGS) and \citet{Einasto:12,Einasto:12a} in a sample of rich clusters in the SDSS.

There is thus a need and appetite in the community for a combination of the targeted studies that can  successfully  identify individual merging events and the large-scale grouping algorithms that can estimate the statistics of clustering  among galaxies. This is the scene in which the present paper is embedded: an attempt to estimate the statistics and characteristics of merging groups of galaxies directly via a grouping algorithm applied to the SDSS.

Throughout this paper we assume the Planck cosmology \citep{2016A&A...594A..13P}: the Hubble constant $H_0 = 67.8~\mathrm{km~s^{-1}Mpc^{-1}}$, the matter density $\Omega_\mathrm{m} = 0.308$, and the dark energy density $\Omega_\Lambda = 0.692$.

\section{Data}

This work is based on catalogue data from the SDSS~DR12 \citep{2011AJ....142...72E, 2015ApJS..219...12A}. We have selected galaxies only from the main contiguous area of the survey (the Legacy Survey). Compared to DR10, which was used for our previous group catalogue \citep{2014A&A...566A...1T}, the main survey area is unchanged, but there have been some slight improvements to the SDSS spectroscopic pipeline.

Galaxy data have been downloaded from the SDSS Catalog Archive Server (CAS\footnote{\url{http://skyserver.sdss3.org/casjobs/}.}). The selection of galaxies and cleaning of the galaxy sample from spurious entries was carried out in the same way as described for the previous catalogue \citep{2014A&A...566A...1T}. Here we list only the main steps:

\begin{enumerate}
\item We selected all objects with the spectroscopic class GALAXY or QSO. The selection was then matched with the photometric data, and only  objects that matched the photometric class GALAXY were kept. The final selection of QSO objects was made manually at a later stage;

\item We visually checked about one thousand bright galaxies  using the SDSS Image List Tool\footnote{\url{http://skyserver.sdss3.org/dr10/en/tools/chart/listinfo.aspx}.}. Based on the inspection, we removed spurious entries;

\item We then filtered out galaxies with the Galactic-extinction-corrected \citep[based on][]{1998ApJ...500..525S} Petrosian $r$-band magnitude fainter than 17.77. The SDSS is incomplete at fainter magnitudes \citep{2002AJ....124.1810S}. After correcting redshift for the motion with respect to the cosmic microwave background\footnote{For CMB correction we used the simplified formula $z_\mathrm{CMB} = z_\mathrm{obs} - v_\mathrm{p}/c$, where $v_\mathrm{p}$ is a motion along the line of sight relative to the CMB. The difference from the correct formula \citep[see e.g.][]{2014MNRAS.442.1117D} $1+z_\mathrm{CMB} = (1+z_\mathrm{obs})/(1+v_\mathrm{p}/c)$ is less than 1~Mpc at the farthest distances.} (CMB), we set the upper distance limit at $z = 0.2$;

\item We complemented the SDSS spectroscopic sample with 1349 redshifts from our previous group catalogue. The added redshifts originated from the Two-degree Field Galaxy Redshift Survey (2dFGRS), the Two Micron All Sky Survey Redshift Survey (2MRS), and the Third Reference Catalogue of Bright Galaxies (RC3). See \citet{2014A&A...566A...1T} for more details.
\end{enumerate}
The final galaxy sample contains 584\,449 entries.

\section{Group finder and detection of merging groups}

\subsection{Friends-of-friends group finder}

The group finder in this work is based on the friends-of-friends (FoF) algorithm, first described by \citet{1976ApJS...32..409T}, \citet{1982ApJ...257..423H}, \citet{1982Natur.300..407Z}, and \citet{1985ApJ...295..368B}. The method works by linking all neighbours of a galaxy within a certain radius, the linking length, to the same system. Previously, we  used the FoF method to detect galaxy groups from the SDSS redshift-space catalogues \citep{2008A&A...479..927T, 2010A&A...514A.102T, 2012A&A...540A.106T, 2014A&A...566A...1T} and from the combined 2MRS, CosmicFlows-2, and 2M++ data sets \citep{2016A&A...588A..14T}.

To scale the linking length with distance, we used the same workflow as described by \citet{2014A&A...566A...1T}. The linking length as a function of $z$ is expressed by an arctan law \begin{equation} 
        d_{LL}(z) = d_{LL,0}\left[ 1+a\arctan(z/z_\star) \right],
        \label{eq:atan}
\end{equation}
where $d_{LL,0}$ is the linking length at $z=0$, and $a$ and $z_\star$ are free parameters. By fitting Eq.~(\ref{eq:atan}) to the linking length scaling relation, we found the following values: $d_{LL,0} = 0.34~\mathrm{Mpc}$, $a=1.4$, and $z_\star=0.09$.

In this work we chose the ratio of the radial to the transversal linking length $b_{||}/b_{\perp}=12$. This value is greater than $b_{||}/b_{\perp}=10$, which we used for our previous SDSS group catalogues, but is the same as in \citet{2016A&A...588A..14T}. The use of a slightly larger value is justified since we apply a membership refinement to the FoF groups (see next section). The use of a higher value is also suggested by \citet{2014MNRAS.440.1763D}, who  analysed the performance of the FoF algorithm on mock catalogues.

\subsection{Group membership refinement}
\label{sect:refine}

The FoF algorithm has a limitation that it can mistake merging groups or groups lying close to each other for a single system.  Nearby field galaxies or filaments connected to groups may also be considered as group members by the FoF method.

\citet{2016A&A...588A..14T} proposed group membership refinement in two steps. First, they used multimodality analysis to split multiple components of groups into separate systems. The second step involved estimation of the virial radius and the escape velocity to exclude group members that are not physically bound to systems. Using the 2MRS data \citet{2016A&A...588A..14T} show that galaxy groups using the FoF method with refinement are in good agreement with galaxy groups detected using the  halo-based approach by \citet{Tully:15}.

In this work we carried out group membership refinement, as described in detail by \citet{2016A&A...588A..14T}. Here we give a brief summary of the steps involved. Multimodality of FoF groups was checked with a model-based clustering analysis, implemented in the statistical computing environment \emph{R}\footnote{\url{https://www.r-project.org}.} in the package \emph{mclust}. We modified the clustering analysis by fixing one coordinate axis with the line of sight, while two other axes were set perpendicular to the line of sight and each other, allowing free orientation in the sky plane. The \emph{mclust} analysis was applied only to groups with at least seven galaxies. For each given number of subgroups (from one to ten), \emph{mclust} gave the most probable locations, sizes, and shapes of subgroups. The number of subgroups was chosen using the Bayesian information criterion. Eventually, each galaxy was assigned to a single group based on the highest probability given by \emph{mclust}.

In the following step assuming the NFW profile we estimated the group virial radius, $R_{200}$, as the radius of a sphere in which the mean matter density is 200 times higher than the mean of the Universe. We excluded a galaxy from a group if the distance in the sky plane between the galaxy and the group centre was greater than the group virial radius. We also excluded a galaxy from its group if its velocity relative to the group centre was higher than the escape velocity at its sky-projected distance from the group centre. The exclusion of galaxies from groups was carried out iteratively. In most cases the refinement converged after a few iterations. This refinement step was only applied  to groups with at least five members.

Finally, we reiterated the group detection and membership refinement procedure on the excluded members in order to determine whether the excluded galaxies formed separate groups. This approach helped us detect small groups that went undetected during multimodality analysis.

\begin{figure}
        \centering
        \includegraphics[width=88mm]{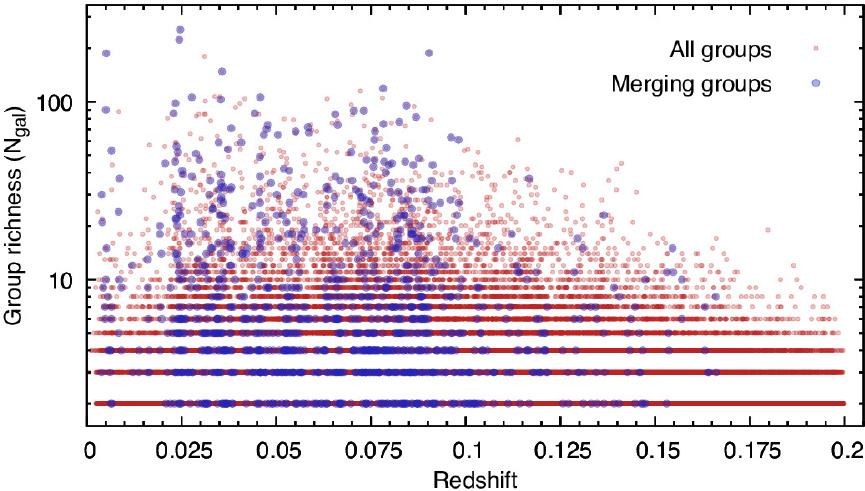}
        \caption{Number of members in a group as a function of redshift. Smaller red points show all groups, while larger blue points indicate potentially merging groups.}
        \label{fig:z_vs_ngal}
\end{figure}

\begin{figure}
        \centering
        \includegraphics[width=88mm]{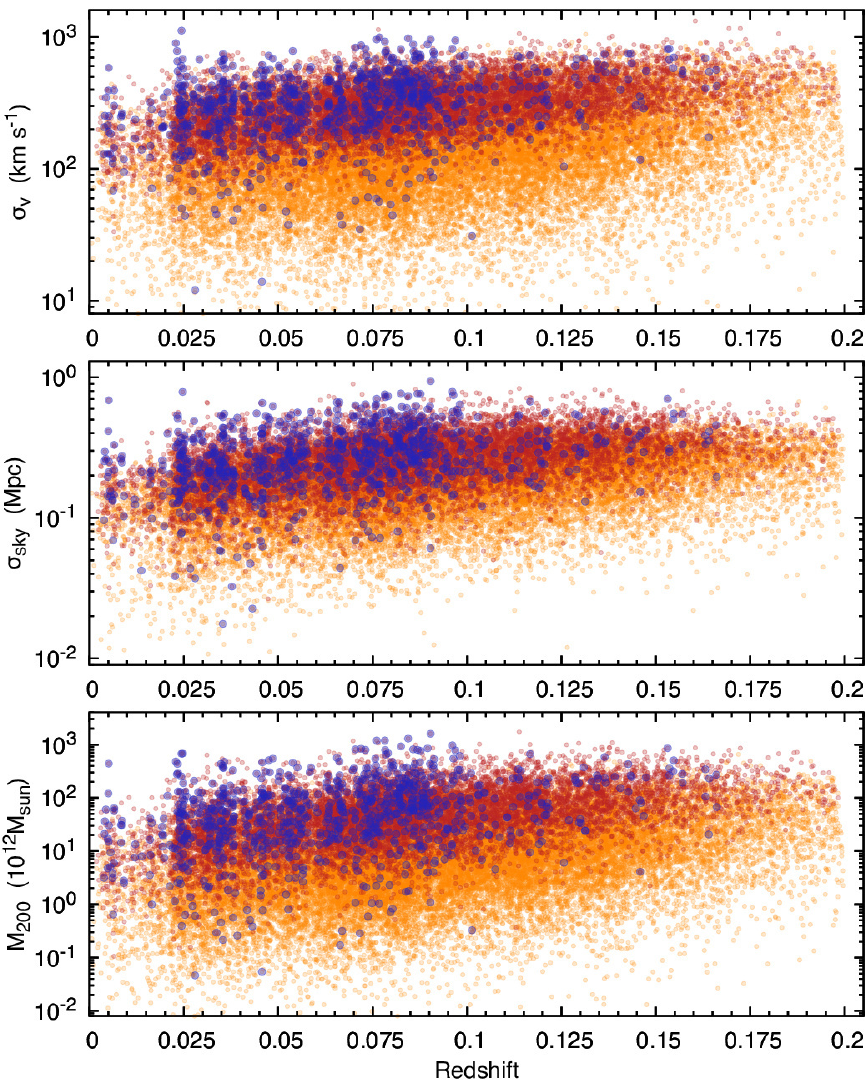}
        \caption{Group velocity dispersion along the line of sight (upper panel), group extent on the sky (middle panel), and group virial mass (lower panel) as a function of redshift. Galaxy pairs are excluded from this figure. Groups with three or four members are shown as orange points, groups with at least five members  as red points, and  potentially merging systems in our catalogue as blue points. Despite the flux-limited nature of our galaxy sample, the group properties are roughly constant with redshift.}
        \label{fig:cl_props}
\end{figure}

\begin{figure*}
        \centering
        \includegraphics[width=180mm]{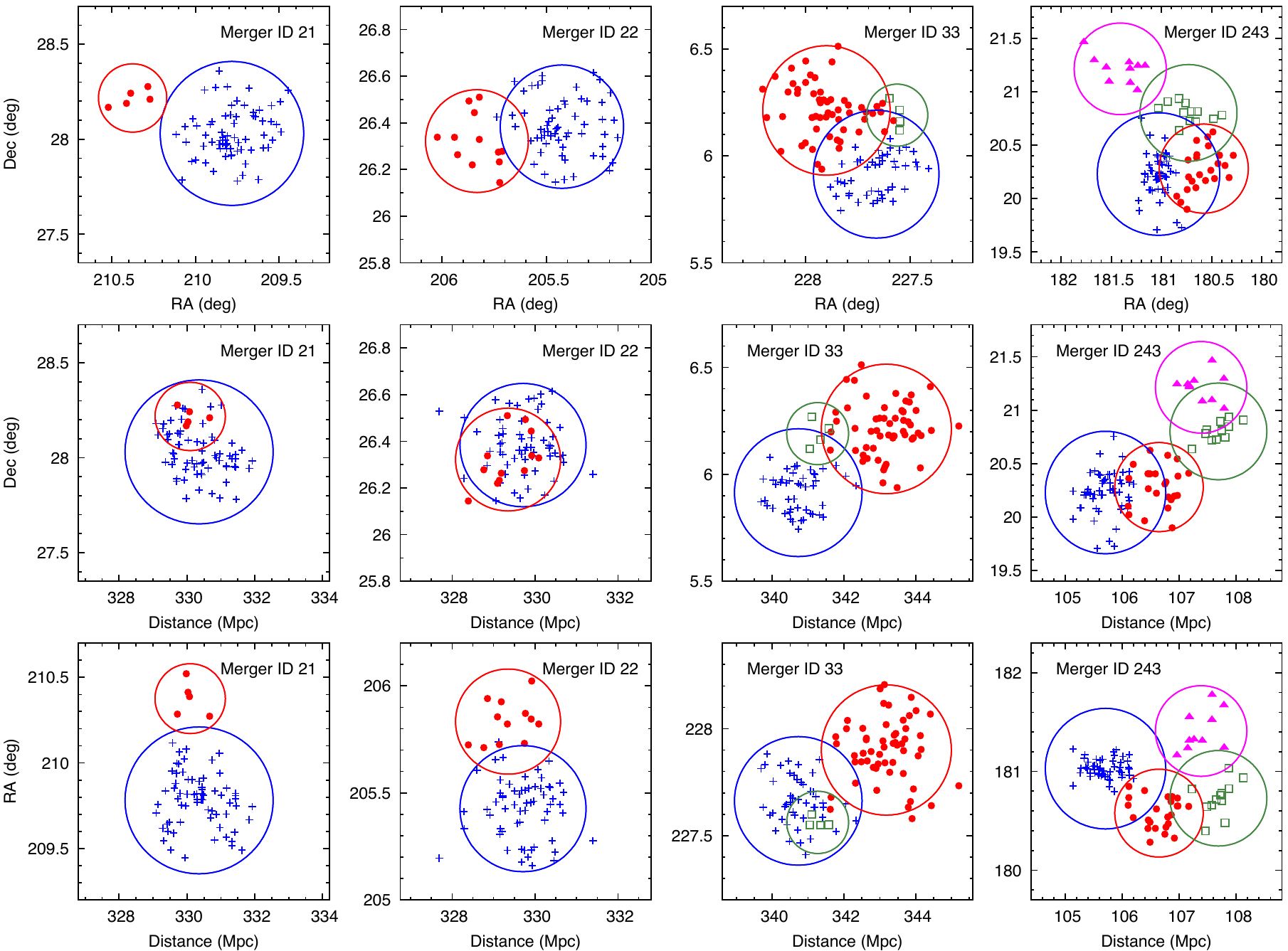}
        \caption{Example of four merging systems in the catalogue. All subgroups in the merging systems are marked with different colours and symbols. The circle around each component shows the estimated size of virial radius $R_{200}$. The scale of the sky coordinates (RA, Dec) are chosen to correspond to the distance scale used in the figure. The top row gives the view in sky plane, while the middle and bottom rows give the view in declination--distance and RA--distance planes, respectively. For distances we use the Fingers-of-God corrected distances which corrects for the smearing of galaxy groups along the line of sight.}
        \label{fig:mergers}
\end{figure*}

\subsection{Detection of potentially merging groups}
\label{subsec:det.mergers}

We consider two galaxy groups in interaction if the distance between their centres (in comoving coordinates) is smaller than the sum of their radii. The group radius in this context is expressed as
\begin{equation}
R_{\mathrm{group}} = 
\begin{cases}
R_{\mathrm{max}},& R_{200} < R_{\mathrm{max}}\\
R_{\mathrm{200}},& R_{\mathrm{max}} \leq R_{200} \leq 2R_{\mathrm{max}}\\
2R_{\mathrm{max}},& R_{200} > 2R_{\mathrm{max}}
\end{cases}
\label{eq:merge_radius}
,\end{equation}
where $R_\mathrm{max}$ is the distance between the group centre and the farthest galaxy in the group in the sky plane, and $R_{200}$ is the virial radius of the group as defined in \citet{2014A&A...566A...1T}. The $R_{200}$ estimated using only observed quantities correlates well with the true value as shown in \citet{2014MNRAS.441.1513O, 2015MNRAS.449.1897O}.

The interacting systems defined in this way are not necessarily real merging systems. Some of the potentially merging systems are probably distinct subgroups of one larger system. The aim of the current criteria is to find systems that are potential mergers. The detailed study of their merger status and properties are left for  future analysis.

\section{Catalogue of galaxy groups}

Using the 584\,449 galaxies in the SDSS main region, we found 88\,662 galaxy groups with at least two members and 498 merging systems. The latter are studied in Sects.~\ref{sec:merging_cat} and \ref{sec:selected}.

The descriptions of all the parameters are given in Appendix~\ref{app:cat}. The details of the calculation of galaxy and group parameters are given in \citet{2012A&A...540A.106T, 2014A&A...566A...1T} and are not repeated   here. The basic properties of our catalogue are summarized in following figures.

Figure~\ref{fig:z_vs_ngal} shows the number of galaxies in a group as a function of redshift. As expected, richer groups are absent at greater distances, due to the flux limited survey (see Fig.~\ref{fig:sample}). Up to redshift 0.1 the maximum number of group members is roughly constant, and decreases thereafter. From Fig.~\ref{fig:z_vs_ngal} we also see that the fraction of merging groups depends on group distance and number of members (see Sect.~\ref{sec:merging_cat} for more details).

Figure~\ref{fig:cl_props} shows the velocity dispersion, group extent on the sky, and virial mass as a function of redshift. Properties of groups with fewer than five members (orange points) are very uncertain and should be treated accordingly. Potentially merging groups identified in this study are marked as  blue points; they cover the full range of group properties (excluding poor groups).

Group virial mass $M_{200}$ used for the membership refinement is estimated directly from observed velocity dispersion and from the group extent on the sky as 
\begin{equation}
        M_{200} \propto \sigma_v^2 \cdot \sigma_\mathrm{sky} .
\end{equation}
The normalization of this relation is described in \citet{2014A&A...566A...1T} together with illustrative figures.

\begin{table}
\caption{ Summary of the frequency of the number of components in merging group systems in the catalogue.} \label{tab:merg.cat.param}
\centering
\begin{tabular}{llllllll}
\hline\hline
Number of components &  2 &  3 &  4 &  5 &  6 & Any \\
\hline
Number of catalogue entries & 458 & 33 &  5 &  1 &  1 & 498 \\
\hline
\end{tabular}
\end{table}

\section{Merging group catalogue}
\label{sec:merging_cat}

\begin{figure}
        \sidecaption
        \includegraphics[width=88mm]{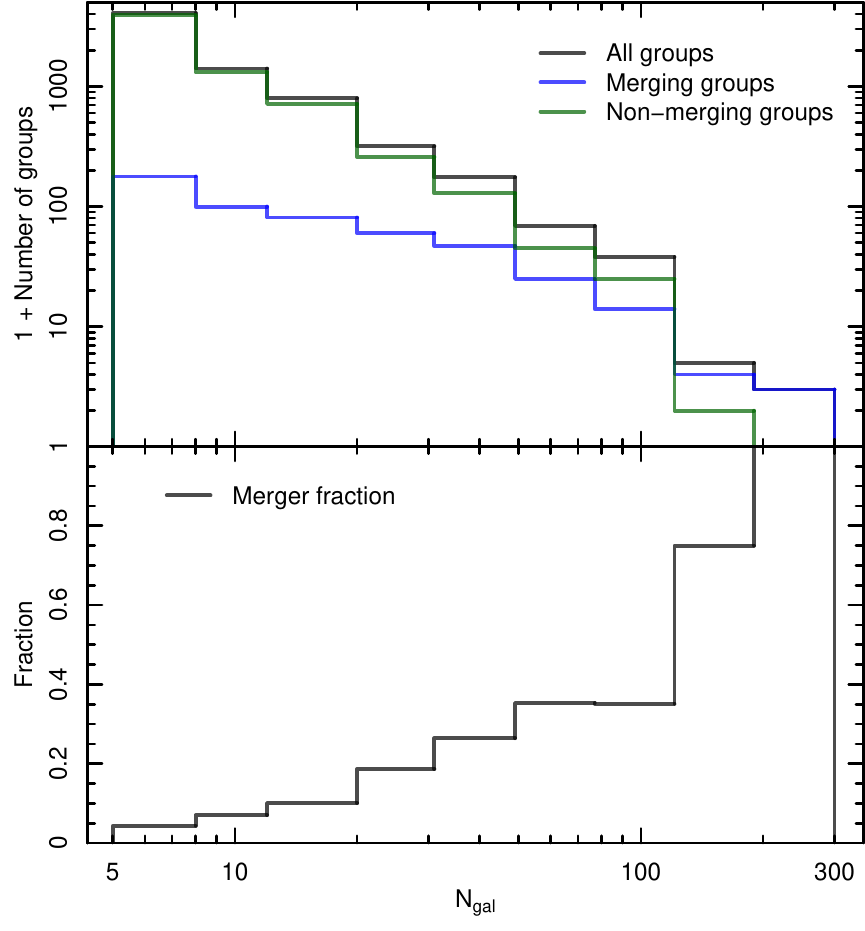}
    \caption{\emph{Upper panel}: Number of groups (black line), mergers (blue line), and non-mergers (green line)  per group richness (number of galaxies in a group) bin.  \emph{Lower panel}: Fraction of potentially merging systems as a function of group richness. The strong increase in merger fraction with group richness in the lower panel is affected by the flux-limited survey.}
    \label{fig:merger.fraction}
\end{figure}

We constructed a catalogue of potentially merging groups by applying the merging group detection algorithm as described in Sect.~\ref{subsec:det.mergers}. In total, the number of potential merger systems in the catalogue is 498,  of which $92\%$ are between two groups. The exact number of entries in the catalogue per number of components is given in Table~\ref{tab:merg.cat.param}. The highest number of components (six) are detected in the Coma cluster, which is studied in Sect.~\ref{sec:selected}. 

Figure~\ref{fig:mergers} shows four examples of potentially merging systems in the catalogue: accretion of smaller component to a cluster (catalogue entry ID~21), the merging  of  two similarly sized components (catalogue entry ID~22), and two multiple mergers (or substructures of elongated clusters; catalogue entries ID~33 and ID~243). We have visually checked all the identified merging systems and visually all of them are very close distinct systems and/or one component is a clear substructure of another component. Some of these systems are potentially interacting systems, which need  to be verified using additional data. More examples and discussion of the merging systems are given in Sect.~\ref{sec:selected}.

By applying our merger definition, we find mergers on very different richness scales (see Fig.~\ref{fig:z_vs_ngal}). The abundance of mergers compared with non-mergers and all groups as a function of number of galaxies in a group is shown in Fig.~\ref{fig:merger.fraction}. The fraction of found mergers is lowest for smaller groups, roughly $5\%$, but increases toward richer groups reaching about $100\%$ at the highest richness end. This does not necessarily mean that all rich clusters are merging systems, it can also mean that they contain substructures. Merging is a dynamical state that should be verified independently. Our analysis only suggests that most rich clusters in our sample are potentially merging systems or contain clearly distinguishable substructures. 

\begin{figure}
        \includegraphics[width=88mm]{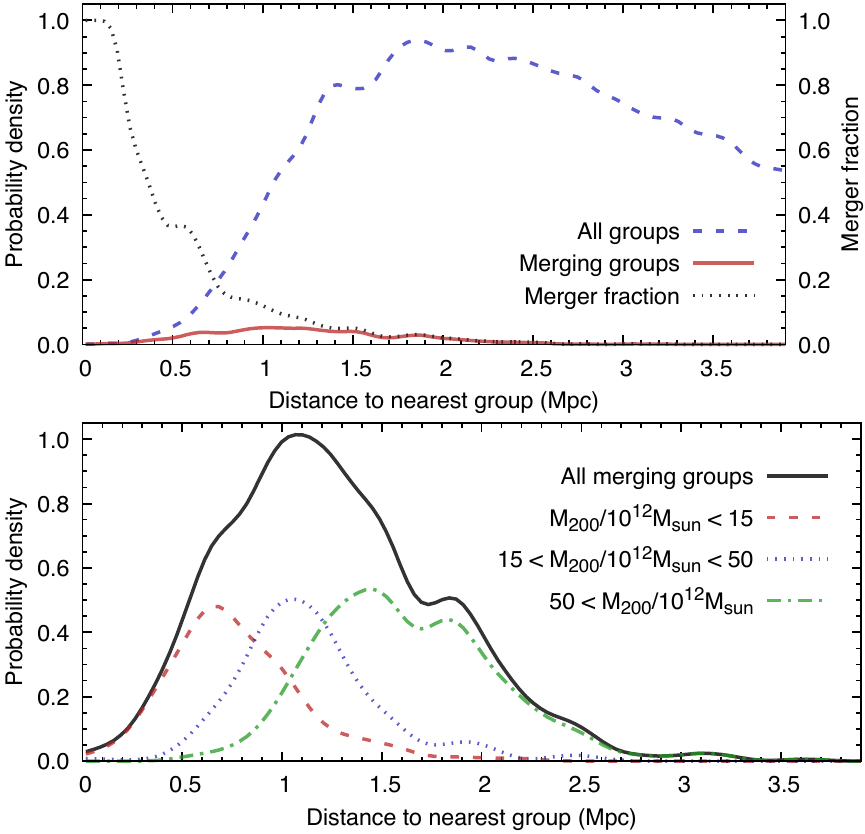}
        \caption{ Distribution of distances to nearest groups/clusters.  \emph{Upper panel}: Nearest group distance distribution for all groups (dashed line) and for merging groups (solid line). The dotted line shows the fraction of merging systems as a function to the nearest group.  \emph{Lower panel}: Nearest distance distribution for merging groups divided by their virial mass $M_{200}$ into three roughly equally sized samples. Since the majority of the merging systems are closer than redshift 0.1, the distributions are calculated for those systems only. When using all clusters, the figure is qualitatively the same.}
        \label{fig:dist.jaotus}
\end{figure} 

A note of caution is due here. The increase in merger fraction as a function of number of group members (see Fig.~\ref{fig:merger.fraction}) is due to the flux-limited survey and the method we use for merger identification. In general, when more data are available,  the probability of detecting mergers/substructure increases regardless of the  method used. From a physical point of view, a more interesting question is whether more massive systems have a higher merger fraction. We addressed this question as follows. We fixed the number of galaxies in a group and studied the merger fraction as a function of group mass. Since the number of groups in a narrow richness bin is relatively low, we did not find any reliable trend with group mass. Hence, the current data are not sufficient to conclude whether more massive groups have a higher fraction of mergers than less massive groups with the same number of galaxies.

Physically speaking, the probability that a system is a merger event increases if the distance between the components is smaller.  The lower panel in Fig.~\ref{fig:dist.jaotus} shows the distribution of  distances between the components for two-component merging systems; it also shows the distance distribution for clusters with different masses. Figure~\ref{fig:dist.jaotus} shows that the distance distribution for more massive clusters is shifted compared with smaller mass clusters, meaning that massive clusters tend to be farther away from each other. This is true in part  because the  method used in this paper does not  distinguish very nearby clusters, and because  the method  can only find clearly distinguishable groups. However, it can also be a physical effect. If a smaller system is merging with a massive cluster, then the smaller group dissolves inside the larger cluster and we do not expect to find subsystems close to the centre of the clusters.

Our definition of potentially merging systems (see Sect.~\ref{subsec:det.mergers}) is rather conservative. The upper panel in Fig.~\ref{fig:dist.jaotus} shows the distribution of distances to the nearest group. The most frequent distance between two groups is around 2~Mpc, decreasing rapidly toward smaller distances. If we look at the nearest group distance distribution for merging systems, we notice that it has a maximum of around 1~Mpc, and there are  very few merging systems identified where the distance is larger than 2~Mpc. Figure~\ref{fig:dist.jaotus} also shows that the fraction of merging systems is 100\% for the smallest distances, and drops rapidly as the distance between groups increases.

\begin{figure}
        \includegraphics[width=88mm]{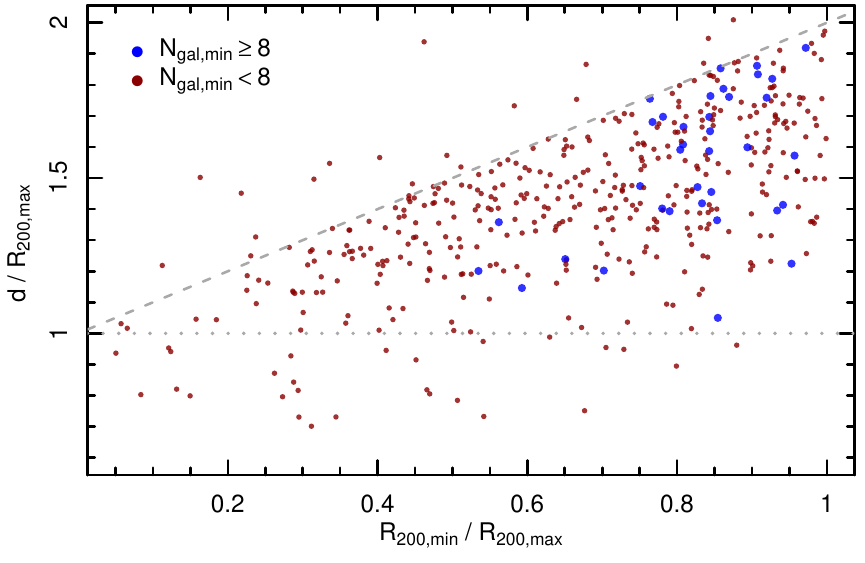}
        \caption{Relative distance (in units of the larger component $R_{200}$) between two component merging systems as a function of their virial radii ratio. Blue points indicate systems where both components contain at least 8 galaxies. The dashed line shows the merging criteria limit for our systems. Some of the smaller systems are above this limit, due to the poorly determined $R_{200}$ value, in which case we have also taken into account the size of the groups in the plane of the sky (see Sect.~\ref{subsec:det.mergers}). The dotted line marks the $R_{200}$ for the larger component. For most of the systems the distance between them is larger than the estimated virial radius of the larger component.}
        \label{fig:dist.in.Rvir}
\end{figure}

Figure~\ref{fig:dist.in.Rvir} shows the relative distance (in units of $R_{200}$) dependence on the size ratio of the components, indicating that the distance between the components does not depend much on the relative size of the components and therefore the method works with the same efficiency regardless of the group size. The underpopulated top left region in the figure is caused by the merging cluster criteria (see Sect.~\ref{subsec:det.mergers}); the outliers in this region are caused by the $R_\mathrm{max}$ part of the definition of $R_\mathrm{group}$ (see Eq.~\ref{eq:merge_radius}). Figure~\ref{fig:dist.in.Rvir} also shows that for most of the merging systems the distance between them is greater than the virial radius ($R_{200}$) of the larger component. There are only a few relatively poor systems where the distance is smaller than the $R_{200}$. This indicates that each individual group is well separated from others and our post processing of FoF groups (see Sect.~\ref{sect:refine}) does not fragment the FoF groups too much.

\begin{figure}
        \includegraphics[width=88mm]{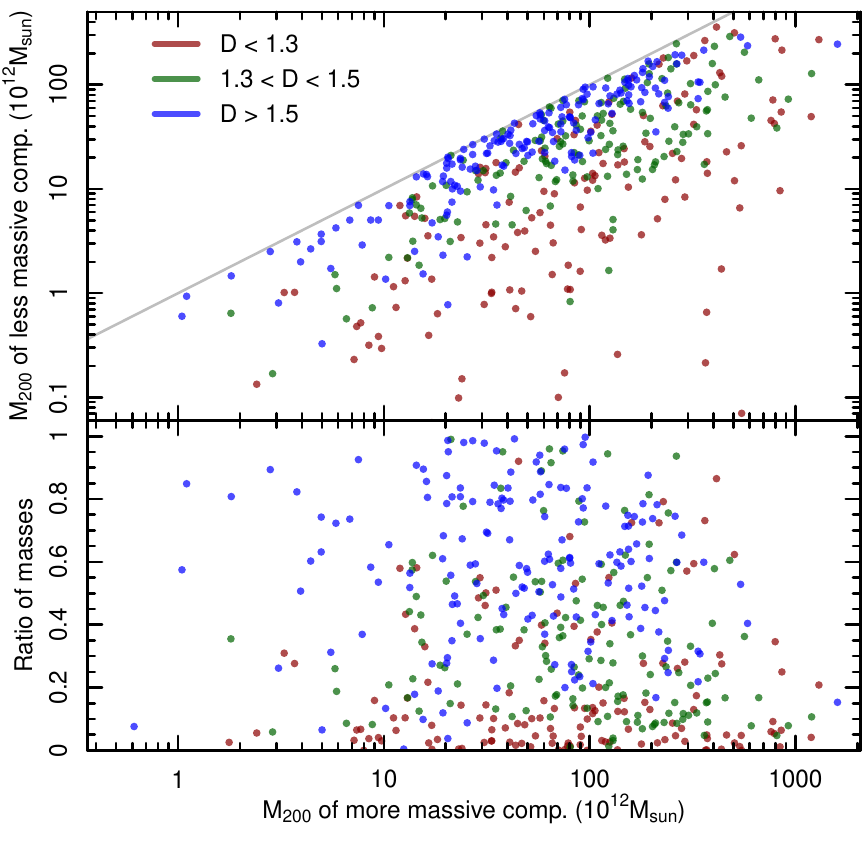}
        \caption{Comparison of the masses of the two-component merging groups. In the top panel the more massive component is shown on the $x$-axis and the less massive one on the $y$-axis. The sample is divided into three subsamples based on the distance between components, normalized with the $R_{200}$: $D = d/\max(R_{200,1}, R_{200,2})$, where $d$ is the distance between  two components. The one-to-one relationship is denoted with a grey line. The bottom panel shows the ratio of the masses as a function of the more massive component.}
        \label{fig:mass_mass}
\end{figure}

As we mentioned earlier, the criteria we use to find potentially merging systems were chosen arbitrarily. The total number of potentially merging systems in the catalogue is rather low. Since the merging system criteria depend only on the distance between galaxy groups, they can be easily redefined to find more potentially merging systems. From Fig.~\ref{fig:dist.jaotus} we see that there are many systems where the distance between groups is less than 1~Mpc that are currently not classified as mergers in the catalogue. In our group catalogue (see Appendix~\ref{app:cat}) we also give the distance to the nearest group that can be independently used to find potentially interacting systems.

\begin{figure*}
        \centering
        \includegraphics[width=180mm]{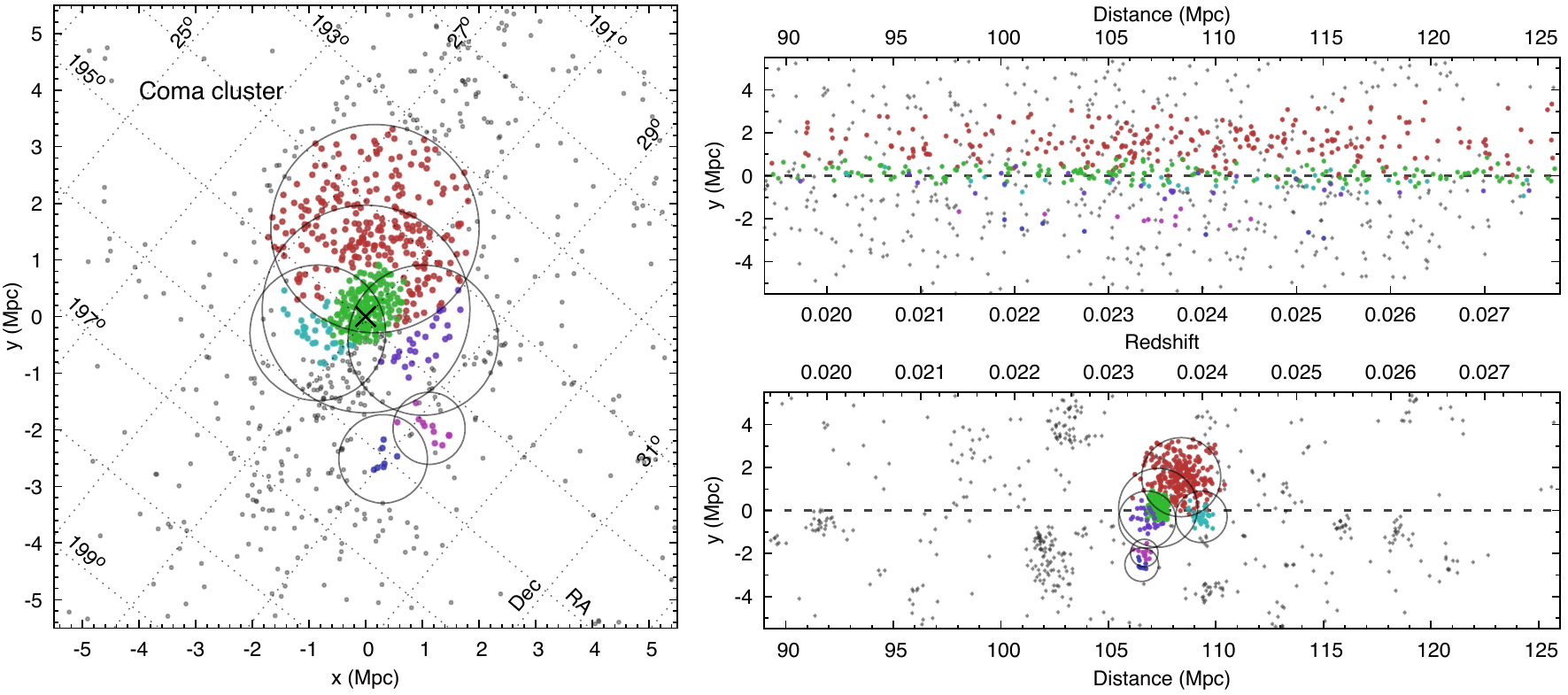}
        \caption{Distribution of galaxies around the Coma cluster region. \emph{Left panel}: Distribution of galaxies in the sky plane (in arbitrary cartesian coordinates), while the right ascensions and declinations are shown as dotted lines. \emph{Upper right panel}: Observed distribution of galaxies along the line of sight. \emph{Lower right panel}: Distribution of galaxies after correction for the redshift space distortions. The redshifts shown in the panels are heliocentric observed redshifts, while the distances are comoving distances given in a CMB frame. Each subcomponent of the Coma cluster is marked as a coloured point and all other galaxies are marked as grey points. The circle around each component is the virial radius ($R_{200}$) of the system.}
        \label{fig:example_coma}
\end{figure*}

\begin{figure*}
        \centering
        \includegraphics[width=174mm]{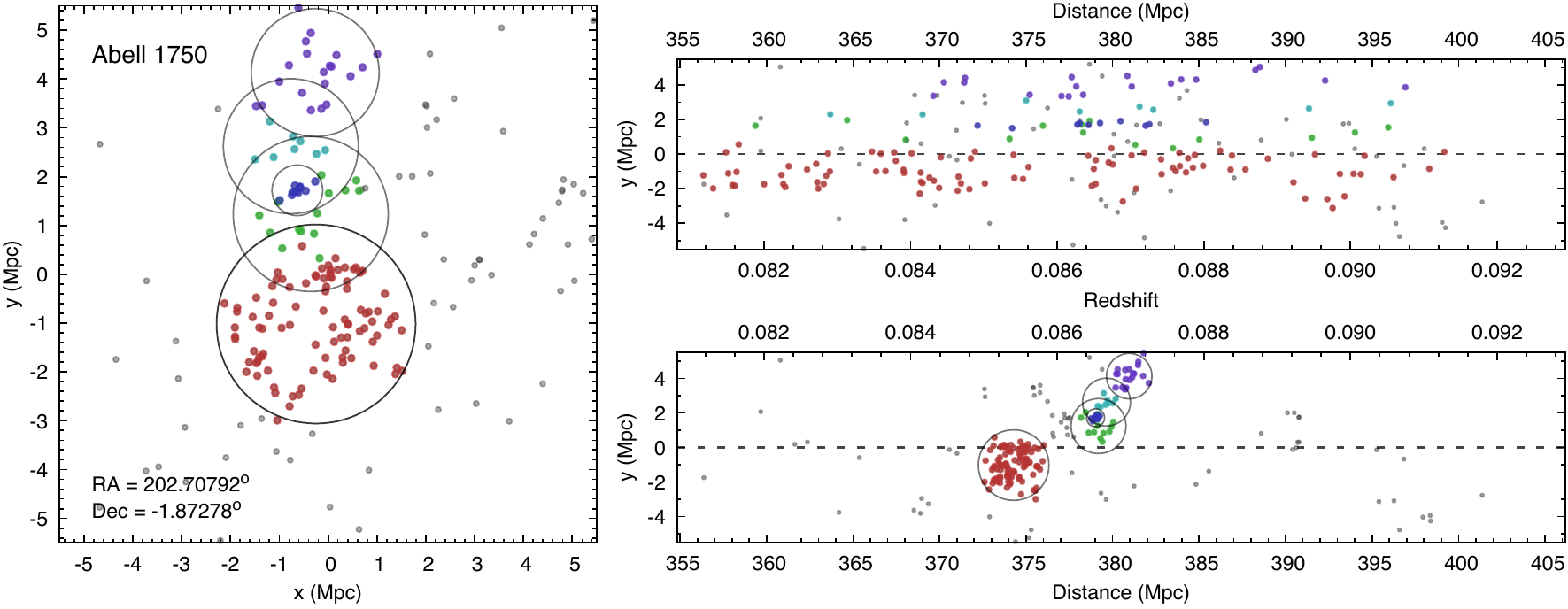}
        \includegraphics[width=174mm]{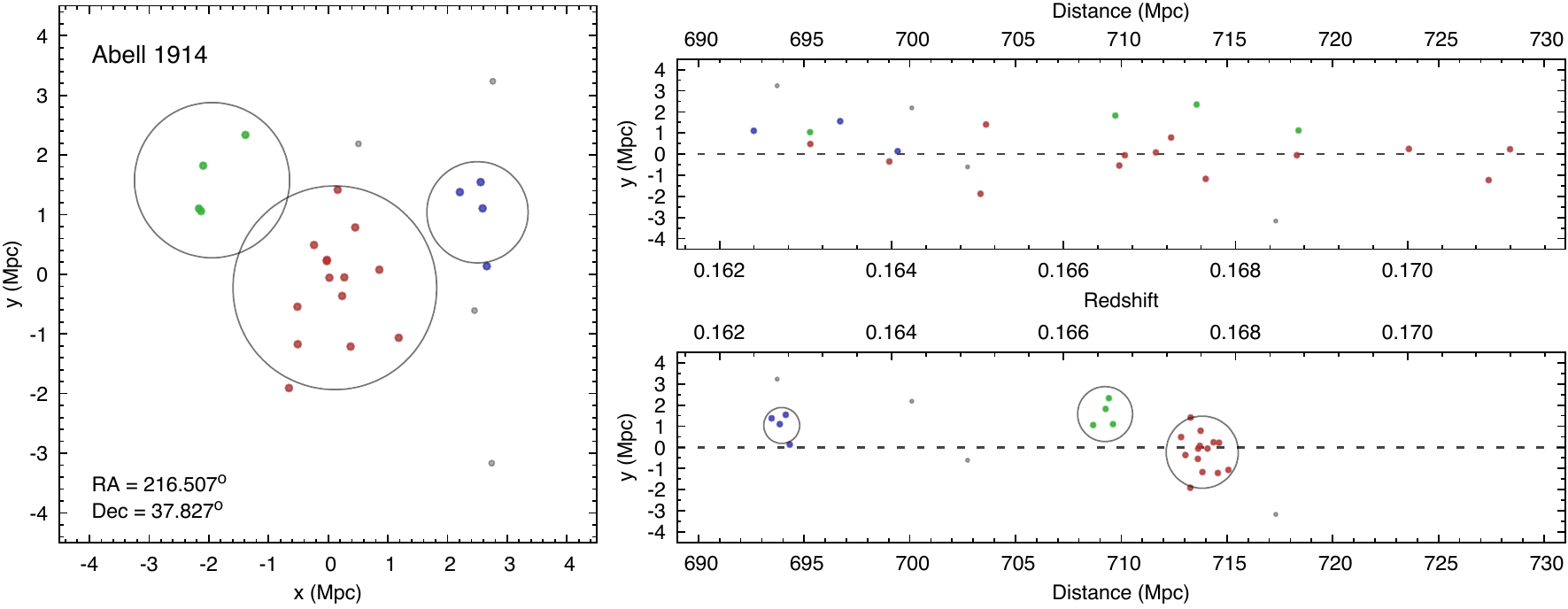}
        \includegraphics[width=174mm]{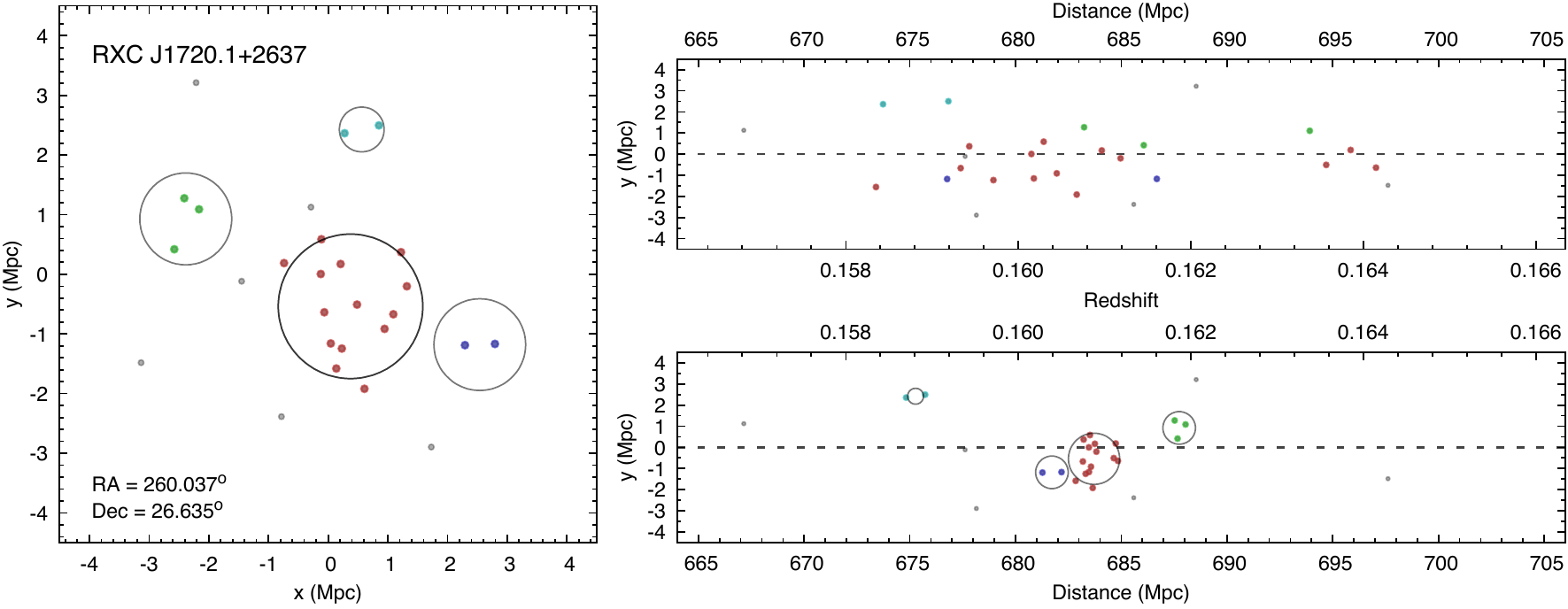}
        \caption{Distribution of galaxies in clusters Abell~1750 (upper panel), Abell~1914 (middle panel), and RXC~J1720.1+2637 (lower panel). The main groups around these systems detected in our catalogue are marked in different colours. The right ascension and declination of the centre of the images are marked in the lower left corner of the left-hand  panels. See caption of Fig.~\ref{fig:example_coma} for detailed description.}
        \label{fig:example_clusters}
\end{figure*}

\subsection{Further taxonomy}

A substantial amount of scientific interest in merging clusters lies in the field of the detection of  dark matter self-interaction  \citep[e.g.][]{2004ApJ...606..819M} and the merging influence on the galaxy star formation/quenching \citep[e.g.][]{2016MNRAS.461.1202J}. For these aspects, it is crucial to know whether the merging systems have passed through each other or not. When  galaxies are described  as point objects, it is difficult to distinguish uniquely these scenarios, and so we do not provide the information on the pre- or post-interaction stage of the mergers.  

Our method does not distinguish the further taxonomy of the merges, e.g. major mergers, accretion events, or substructures. Figure~\ref{fig:mass_mass} shows the relative masses of the two-component merging systems. From the figure it can be seen that the masses of different components are mostly near the one-to-one line, suggesting that a substantial number of mergers might be major ones (like the Bullet Cluster, \citealt{2002ApJ...567L..27M}). Therefore the  further taxonomy of the merges (dividing the cases as mergers, accretion, or substructure with respect to the  visible region) is plausible, but depending on the application, is subjective and therefore deserves a separate study.

\section{Selected known clusters}
\label{sec:selected}

In this section we compare our detected groups with some known clusters in the nearby Universe. The aim of this section is to illustrate the capabilities of our group catalogue and to discuss some aspects of merging cluster identification. A detailed comparison with all known merging clusters\footnote{For examples of the clusters identified by the Merging Cluster Collaboration: \url{http://www.mergingclustercollaboration.org}.} in the SDSS region is beyond the scope of the  current paper.

In Fig.~\ref{fig:example_coma} we show the galaxy distribution in the \object{Coma cluster} region. In our group catalogue, the Coma cluster belongs to a merging system with six subsystems (marked with various colours in Fig.~\ref{fig:example_coma}). The centre region of the Coma cluster (green points) is a clearly distinguishable component in the Coma cluster. The centre of this component is located at redshift 0.0235, which is slightly higher than the value (0.0231) given in the NASA/IPAC Extragalactic Database\footnote{\url{http://ned.ipac.caltech.edu}.} \citep{1999ApJS..125...35S}, but is consistent with the value given in \citet{2017ApJS..229...20S}. The difference comes from the identification of the Coma cluster members. Around the dense core of the Coma cluster are three other subsystems that form the outskirts of the Coma cluster. In our merging system catalogue two smaller groups are also connected with the Coma cluster that probably belong to a filament connected with it. The substructure of the Coma cluster is rather rich and several subcomponents can be identified \citep{Fitchett:87,Gambera:97,Adami:05,Adami:09}.

Figure~\ref{fig:example_coma} (see also Fig.~\ref{fig:mergers}) shows that the virial radius of groups ($R_{200}$) is a good representative of a cluster scale. The only exception is the core region (green points) where the $R_{200}$ is larger than the visible extent of the cluster. Since the $R_{200}$ is directly derived from $M_{200}$, which most strongly depends on the velocity dispersion, the large virial radius for the core region is attributed to the large velocity dispersion along the line of sight. Most likely, the transition from the core region to the outskirts  is smooth and the classification of galaxies between distinct components is not physically motivated. We note that for complicated merging systems, the inner structure of the systems should be clarified visually and with the help of additional data.

The next cluster we chose is \object{Abell 1750}, which is a strongly merging double cluster \citep[see e.g.][]{Belsole:04, Einasto:10, 2016ApJ...818..131B}. The distribution of galaxies in this cluster is shown in the upper panel of Fig.~\ref{fig:example_clusters}. The main component of this cluster is not identified as a merging system based on our criteria. However, next to the main component lies a four-component merging system. A visual check confirms that this is actually connected with the main body of Abell~1750, and indeed Abell~1750 is a merging system. The non-identification as a merging system in our catalogue is due to the criteria we used. Using  slightly less strict criteria, the Abell~1750 becomes a merging system in our data.

The sky distribution of galaxies around Abell~1750 indicates that there is a filamentary infall of galaxies towards the cluster. The Fingers-of-God corrected distribution of galaxies (see Fig.~\ref{fig:example_clusters}) in the Abell~1750 region shows clearly that there is a filament that is in connection with the main body of the cluster. This filamentary structure is not visible in the redshift distribution, but becomes distinct when the galaxy positions are corrected for the random motions inside galaxy groups.

The middle and lower panels in Fig.~\ref{fig:example_clusters} show the merging systems \object{Abell 1914} and \object{RXC J1720.1+2637} \citep{2012MNRAS.420.2120M}. Both clusters are located at around redshift 0.16--0.17. Owing to the relatively large distances these systems contain significantly fewer galaxies than the two previous examples. These systems are not identified as merging systems in our catalogue, but visual inspection indicates that around the main body of the cluster there are other smaller systems that might  potentially be merging systems. 

A small sample of merging clusters has  recently been constructed by \citet{deLosRios:16} using simulations and machine learning to train their algorithm. Among the systems in their sample there are two mergers  that are in the same SDSS region that we analyse, namely clusters \object{Abell 1424} and \object{Abell 2142}. In the case of Abell~1424, our method detected three groups in this region, which are not part of our merger catalogue;  the distance between individual groups is higher than the sum of virial radii (merging criteria in our catalogue). In the case of Abell~2142 we detected this as a merging system, but with different components due to the lack of clear separation between subcomponents.

Based on the previous examples we can conclude that the merging system definition in this paper is rather conservative and that there are merging systems that are not marked as mergers in our catalogue. However, this definition can  easily be adjusted by increasing the allowed distance between merging components.

The right-hand panels of Figs.~\ref{fig:example_coma} and \ref{fig:example_clusters} show the effect of the Fingers-of-God suppression\footnote{The redshift-space distortions of groups are suppressed using the velocity dispersion along the line of sight and the group extent on the sky plane. See \citet{2012A&A...539A..80L} and \citet{2014A&A...566A...1T} for a detailed description of this procedure.}. In the upper panels we show the observed redshift distribution of the galaxies, while in the  lower panel the corrected distribution of galaxies is shown. These figures clearly highlight how the chaotic motions in galaxy groups smear out the structure along the line of sight. However, after the redshift-space corrections, the smearing effect of Fingers-of-God is suppressed and the large-scale structure around the galaxy clusters is visible. Here we emphasize that the correction is only statistical and it cannot be used to assess the inner structure of galaxy groups along the line of sight. The corrected positions of galaxies can only be used for the large-scale structure studies such as supercluster detection \citep{2012A&A...539A..80L} or filament detection \citep{2014MNRAS.438.3465T}.

\section{Mass-to-light ratio of galaxy groups}

\begin{figure}
        \centering
        \includegraphics[width=88mm]{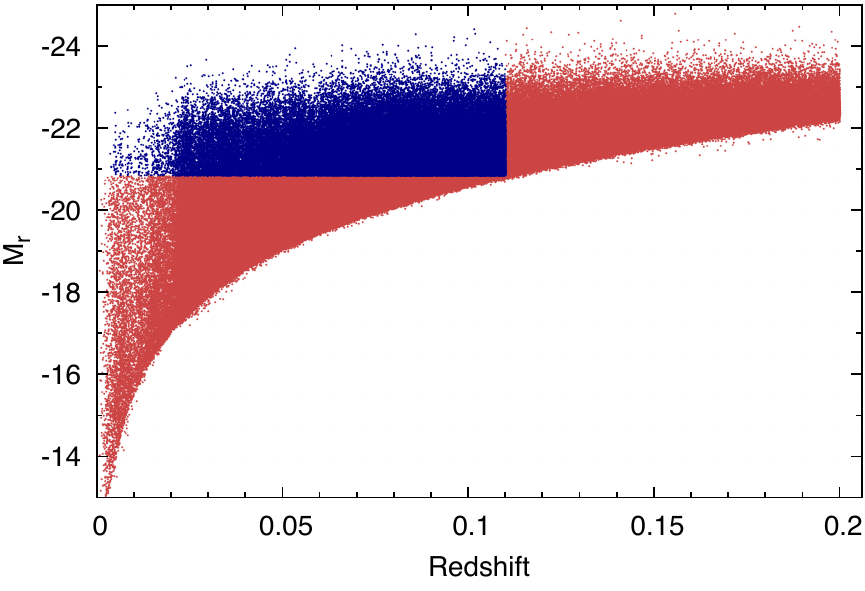}
        \caption{Absolute $r$ magnitudes of galaxies as a function of redshift. Red points denote all galaxies in the flux-limited sample. Blue points denote galaxies belonging to groups in the volume-limited sample.}
        \label{fig:sample}
\end{figure}

In this section we study the $M/L$ of galaxy groups and the relation of the $M/L$ to group mass and richness. To avoid selection effects, we created a volume-limited sample of galaxies with the absolute $r$ magnitude ($M_r$) brighter than $-20.84$, corresponding to the following condition: $M_r+5\log h<-20.0$\footnote{Here the Hubble constant is represented as $H_0 = 100\,h\,\mathrm{km}\,\mathrm{s}^{-1}\mathrm{Mpc}^{-1}$. This limit was chosen based on  \citet{2014A&A...566A...1T}.}. The sample is complete up to the redshift of 0.11. We applied the redshift limit to the group mean redshift rather than the redshift of individual galaxies. Figure~\ref{fig:sample} shows the original flux-limited sample of galaxies together with the volume-limited sample. We then recalculated the group richness ($N_\mathrm{gal}$) and total $r$-band luminosity ($L_r$) of groups, based on galaxies in the volume-limited sample. A weight of 1.858 was applied to the luminosity $L_r$, accounting for the missing light from fainter galaxies \citep[see][]{2011A&A...529A..53T, 2014A&A...566A...1T}. Group masses ($M_{200}$), computed within virial radii $R_{200}$, where the mass density exceeds 200 times the critical density of the Universe, were adopted from the group catalogue.

\begin{figure}
        \centering
        \includegraphics[width=88mm]{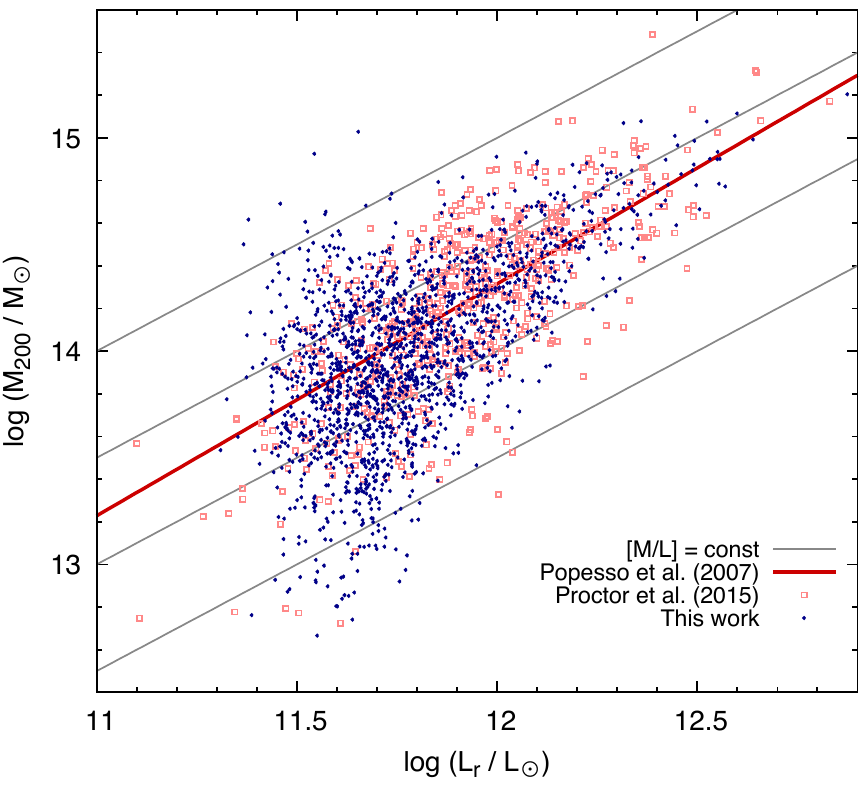}
        \caption{Masses of galaxy groups as a function of group total $r$-band luminosity. Blue points denote our data in the volume-limited sample with at least six group members. The red open squares and the red line indicate data from the literature \citep{2007A&A...464..451P, 2015MNRAS.449.2345P}. The grey lines show constant $\log(M/L)$ values of 3.0, 2.5, 2.0, and 1.5 (from top left to bottom right).}
        \label{fig:ML_proctor}
\end{figure}

Analogously to the volume-limited sample of galaxies, we selected galaxy groups with $z_\mathrm{group}<0.11$ and we further restricted the sample to groups with $N_\mathrm{gal} \geq 6$ as poorer groups have unreliable mass estimates. This left us with 1716 groups, covering a mass range of ${\sim}10^{12}\ldots 10^{15} M_{\odot}$.

In Fig.~\ref{fig:ML_proctor}, masses of galaxy groups are plotted against luminosity $L_r$. For comparison, we have also plotted group properties from the sample of the maxBCG catalogue \citep{2015MNRAS.449.2345P} and the mass-luminosity relation by \citet{2007A&A...464..451P}. It can be seen from the figure that our data agree well with the literature data. Compared to the maxBCG sample, our sample contains a larger number of poor groups, which show higher scatter in mass.  \citet{2015MNRAS.449.2345P} find that the slope in the mass-luminosity relation increases at lower masses. We do not see a pronounced change in the slope from our data in Fig.~\ref{fig:ML_proctor}.

\begin{figure}
        \centering
        \includegraphics[width=88mm]{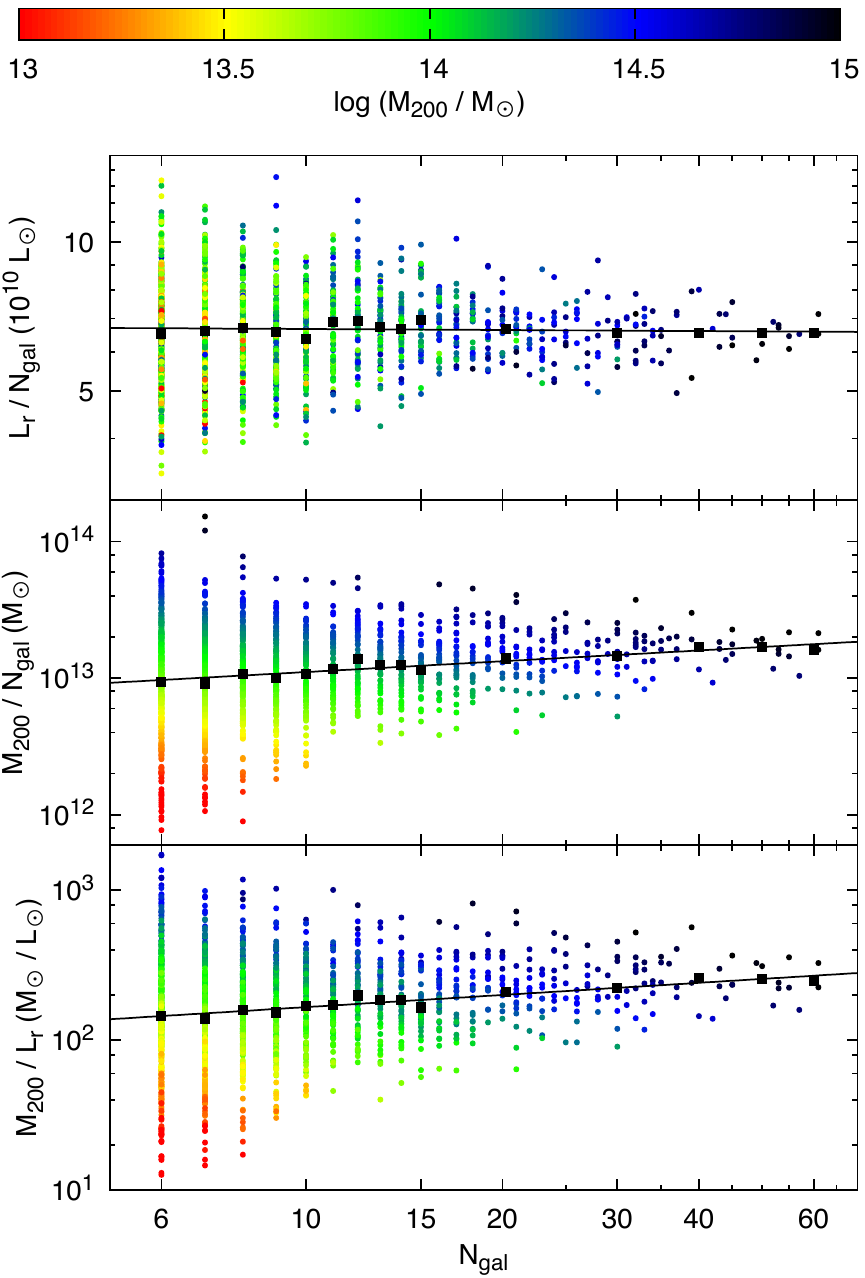}
        \caption{Group luminosity-richness ratio (top panel), mass-richness ratio (middle panel), and mass-to-light ratio (bottom panel) as functions of richness in the volume-limited sample ($M_r<-20.84$). Different group masses are indicated according to the colour bar at the top. Black squares give mean values in richness bins, solid black lines are linear least-square fits to the binned data.}
        \label{fig:groups_vollim}
\end{figure}

The scatter in Fig.~\ref{fig:ML_proctor} is dominated by errors in mass estimation. The increased scatter toward low luminosity groups is caused by the decreasing number of galaxies in groups, which leads to the less accurate mass estimation \citep[see][]{2014MNRAS.441.1513O, 2015MNRAS.449.1897O}. It can also be  deduced from Fig.~\ref{fig:groups_vollim}: for poorer groups the differences between luminosities are about two-three times (upper panel), while for group masses the differences are about two orders of magnitude (middle panel).

Figure~\ref{fig:groups_vollim} shows the $L_r/N_\mathrm{gal}$, $M_{200}/N_\mathrm{gal}$, and $M/L$ as functions of group richness. It can be seen from the top panel of Fig.~\ref{fig:groups_vollim} that the group luminosity-richness ratio is constant over the full richness range (slope $-0.007\pm0.009$), which means that the $L_r$--$N_\mathrm{gal}$ relation is linear. The linear luminosity-richness relation for SDSS galaxies was also found by \citet{2007A&A...464..451P} and \citet{2009ApJS..183..197W}.

The middle panel of Fig.~\ref{fig:groups_vollim} shows that the group mass per member galaxy increases with richness. The relation can be expressed by a power law: $M_{200}/N_\mathrm{gal} \propto N_\mathrm{gal}^{\alpha}$, with $\alpha = 0.26 \pm 0.03$. However, by following groups with the same mass (same colour in the figure), richer groups have lower $M_{200}/N_\mathrm{gal}$ than poorer groups. The same relation holds for $M/L$ (bottom panel of Fig.~\ref{fig:groups_vollim}), due to the proportionality between $L_r$ and $N_\mathrm{gal}$.

To conclude, $M/L$  increases slightly  with group richness. If we look at groups with similar mass, the $M/L$ decreases  with group richness, as logically expected.

\section{Conclusions and discussions}

We used the data from the SDSS DR12 and applied a modified FoF group finding algorithm \citep{2016A&A...588A..14T} in order to detect galaxy groups and clusters in the data. In a modified group finding algorithm the conventional FoF method is complemented with a multimodality clustering analysis to distinguish individual subcomponents in traditional FoF groups. In addition to the common galaxy and group catalogue, we make available the potentially merging system catalogue. See Appendix~\ref{app:cat} for the description of the published products.

The improved FoF group finding algorithm was motivated by the desire to suppress the redshift space distortions of galaxy groups, the so-called Fingers-of-God effect \citep{1972MNRAS.156P...1J, 1978IAUS...79...31T}. For proper corrections and to recover the underlying large-scale structure, single subcomponents of interacting systems should be individually treated. As illustrated in this paper and in \citet{2016A&A...588A..14T}, the implemented redshift-space distortions correction works fairly well. The proper grouping of galaxies is a prerequisite for detecting the large-scale structure filaments that connect galaxy groups and clusters \citep{2014MNRAS.438.3465T, 2016A&C....16...17T}. Additionally, proper grouping is essential in order to reconstruct the matter and velocity field via Wiener Filtering \citep{Sorce:2017}.

To determine the potentially merging systems in our data, we utilize the group virial radii. The merging systems are defined in a simple way, namely the radii of two systems should overlap. We illustrated some of the detected merging systems in our paper and showed that our definition of merging systems is rather conservative. In our catalogue  we calculated the distance of each group to the nearest system, which can be easily used to redefine the merging system criteria for other studies.

Using our selected merging systems, we conclude that due to the complicated nature of merging systems, the automated detection of merging systems is not a straightforward procedure. For the detailed analysis of merging systems, each system should be visually inspected. Our analysis shows that there are known merging systems that were not marked as mergers in our catalogue. Additionally, some of the mergers in our catalogue are just subcomponents of single clusters. This should be taken into account while using the data  for potentially merging systems published here.

For a comparison with known clusters we selected four systems including the Coma cluster and Abell~1750. The Coma cluster in our catalogue is a merging system with six distinguishable  subcomponents; the central component of the Coma cluster is located at redshift 0.0235, which is slightly higher than the  previously referenced value of 0.0231 \citep{1999ApJS..125...35S}. In the case of Abell~1750 we found a clear sign of filamentary infall toward this cluster, which was very clearly visible after the suppression of redshift-space distortions.

We also analysed the $M/L$ of galaxy groups and clusters in our sample. Our analysis shows that the $M/L$ increases when group mass increases. Additionally, the $M/L$ depends on the group richness, being lower for richer groups with the same mass. 
In our $M/L$ analysis, the maximum value reaches 400, which is in good agreement with the value derived by \citet{2014MNRAS.439.2505B}.

In forthcoming studies, the published catalogues will be used for a variety of purposes, including the detection of galaxy filaments and superclusters in the data, and the analysis of galaxy properties in merging systems.

\begin{acknowledgements}
We thank the Referee for the constructive report that helped us to clarify and improve the paper. We thank Boris Deshev and Mirt Gramann for their contribution and useful comments. This work was supported by institutional research funding IUT26-2 and IUT40-2
of the Estonian Ministry of Education and Research. We acknowledge the support
by the Centre of Excellence “Dark side of the Universe” (TK133) financed by the
European Union through the European Regional Development Fund. This research has made use of the NASA’s Astrophysics Data System Bibliographic
Services.
Funding for SDSS-III has been provided by the Alfred P. Sloan Foundation, the
Participating Institutions, the National Science Foundation, and the U.S.
Department of Energy Office of Science. The SDSS-III web site is
http://www.sdss3.org/.
SDSS-III is managed by the Astrophysical Research Consortium for the
Participating Institutions of the SDSS-III Collaboration including the
University of Arizona, the Brazilian Participation Group, Brookhaven National
Laboratory, Carnegie Mellon University, the University of Florida, the French
Participation Group, the German Participation Group, Harvard University, the
Instituto de Astrofisica de Canarias, the Michigan State/Notre Dame/JINA
Participation Group, Johns Hopkins University, Lawrence Berkeley National
Laboratory, Max Planck Institute for Astrophysics, Max Planck Institute for
Extraterrestrial Physics, New Mexico State University, New York University,
Ohio State University, Pennsylvania State University, University of Portsmouth,
Princeton University, the Spanish Participation Group, University of Tokyo, the
University of Utah, Vanderbilt University, the University of Virginia, the University
of Washington, and Yale University.
\end{acknowledgements}

\appendix

\section{Description of group catalogues}
\label{app:cat}

    The catalogue of galaxy groups consists of three tables. The first table lists the galaxies that were used to generate the catalogue of groups, the second  describes the group properties, and the third  lists the merging systems identified in this study. We refer to \citet{2012A&A...540A.106T, 2014A&A...566A...1T, 2016A&A...588A..14T} for a detailed description of all the parameters that are given in the catalogues. The catalogues are available at \url{http://cosmodb.to.ee}. The catalogues will also be made available  through the Strasbourg Astronomical Data Centre (CDS).

\subsection{Description of the galaxy catalogue}

The galaxy catalogue contains the following information (column numbers are given in square brackets):
\begin{enumerate}
    \item{[1]\,\texttt{galid} --} our unique identification number for the galaxies;
    \item{[2]\,\texttt{specobjid} --} SDSS spectroscopic object identification number;
    \item{[3]\,\texttt{objid} --} SDSS photometric object identification number;
    \item{[4]\,\texttt{groupid} --}  group/cluster ID given in the present paper;
    \item{[5]\,\texttt{ngal} --} richness (number of members) of the group/cluster the galaxy belongs to;
    \item{[6]\,\texttt{rank} --} luminosity rank of the galaxy within its group; rank~1 indicates the most luminous galaxy;
    \item{[7]\,\texttt{groupdist} --} comoving distance to the group/cluster centre  the galaxy belongs to, in units of Mpc, calculated as an average over all galaxies within the group/cluster;
    \item{[8]\,\texttt{zobs} --} observed redshift (without the CMB correction);
    \item{[9]\,\texttt{zcmb} --} redshift, corrected to the CMB rest frame;
    \item{[10]\,\texttt{zerr}--} uncertainty of the redshift;
    \item{[11]\,\texttt{dist} --} comoving distance in units of Mpc (calculated directly from the CMB-corrected redshift);
    \item{[12]\,\texttt{dist\_cor} --} comoving distance of the galaxy after suppressing  the Finger-of-God effect;
    \item{[13--14]\,\texttt{raj2000, dej2000} --} right ascension and declination (deg);
    \item{[15--16]\,\texttt{glon, glat} --} Galactic longitude and latitude (deg);
    \item{[17--18]\,\texttt{sglon, sglat} --} supergalactic longitude and latitude (deg);
    \item{[19--20]\,\texttt{lam, eta} --} SDSS survey coordinates $\lambda$ and $\eta$ (deg);
    \item{[21--3]\,\texttt{crd\_xyz} --} cartesian coordinates defined by $\eta$ and $\lambda$;
    \item{[24--28]\,\texttt{mag\_$x$} --} Galactic-extinction-corrected Petrosian magnitude ($x\in ugriz$ filters);
    \item{[29--33]\,\texttt{absmag\_$x$} --} absolute magnitude of the galaxy, $k$+$e$-corrected ($x\in ugriz$ filters);
    \item{[34--38]\,\texttt{kecor\_$x$} --} $k+e$-correction ($x\in ugriz$ filters);
    \item{[39--43]\,\texttt{ext\_$x$}} Galactic extinction ($x\in ugriz$ filters);
    \item{[44]\,\texttt{lum\_r} --} observed luminosity in the $r$ band in units of $10^{10}L_\odot$, where $L_\odot=4.64$\,mag \citep{2007AJ....133..734B};
    \item{[45]\,\texttt{weight} --} weight factor for the galaxy (\texttt{w}$\cdot$\texttt{lum\_r} was used to calculate the luminosity density field);
    \item{[46]\,\texttt{hc\_e} --} probability of being an early-type galaxy \citep[from][]{2011A&A...525A.157H};
    \item{[47]\,\texttt{hc\_s0} --} probability of being an S0 galaxy;
    \item{[48]\,\texttt{hc\_sab} --} probability of being an Sab galaxy;
    \item{[49]\,\texttt{hc\_scd} --} probability of being an Scd galaxy;
    \item{[50]\,\texttt{dist\_edge} --} comoving distance of the galaxy from the border of the survey mask;
    \item{[51--54]\,\texttt{den$a$} --} normalized environmental density of the galaxy for various smoothing scales ($a=1.5,\,3,\,6,\,10$~Mpc).
\end{enumerate}

\subsection{Description of the group/cluster catalogue}

The catalogue of groups/clusters contains the following information (column numbers are given in square brackets):
\begin{enumerate}
    \item{[1]\,\texttt{groupid} --} group/cluster ID;
    \item{[2]\,\texttt{ngal} --} richness (number of members) of the group;
    \item{[3--4]\,\texttt{raj2000, dej2000} --} right ascension and declination (deg) of the group centre;
    \item{[5--6]\,\texttt{lam, eta} --} SDSS survey coordinates $\lambda$ and $\eta$ (deg) of the group centre;
        \item{[7--9]\,\texttt{crd\_xyz} --} cartesian coordinates of the group centre;
        \item{[10]\,\texttt{zcmb} --} CMB-corrected redshift of group, calculated as an average over all group/cluster members;
    \item{[11]\,\texttt{groupdist} --} comoving distance to the group centre (Mpc);
    \item{[12]\,\texttt{sigma\_v} --} rms radial velocity deviation ($\sigma_v$ in physical coordinates,  in \mbox{km\,s$^{-1}$});
    \item{[13]\,\texttt{sigma\_sky} --} rms deviation of the projected distance in the sky from the group centre ($\sigma_\mathrm{sky}$ in physical coordinates, in Mpc), $\sigma_\mathrm{sky}$ defines the extent of the group in the sky;
    \item{[14]\,\texttt{r\_max} --} maximum radius of the group/cluster in the plane of the sky (Mpc);
    \item{[15]\,\texttt{mass\_200} --} estimated mass of the group assuming the NFW density profile (in units of $10^{12}M_\odot$);
    \item{[16]\,\texttt{r\_200} --} radius in which the group mean density is 200 times higher than the average density of the Universe (in units of Mpc); radius of the group that contains the mass $M_{200}$;
    \item{[17]\,\texttt{lum\_r\_group} --} observed luminosity, i.e. the sum of luminosities of the galaxies in the group/cluster ($10^{10}L_\odot$);
    \item{[18]\,\texttt{weight} --} weight factor for the group at the mean distance of the group;
    \item{[19--22]\,\texttt{den$a$} --} normalized environmental density (mean of group galaxy densities) of the group for various smoothing scales ($a=1.5,\,3,\,6,\,10$~Mpc).
        \item{[23]\,\texttt{dist\_nearest\_cl} --} distance to the nearest cluster in comoving coordinates (Mpc);
        \item{[24]\,\texttt{id\_nearest\_cl} --} \texttt{groupid} of the nearest cluster;
        \item{[25]\,\texttt{id\_merger} --} ID of the merging system, zero otherwise.
\end{enumerate}

\subsection{Description of the merging system catalogue}

The catalogue of merging systems contains the following information (column numbers are given in square brackets):
\begin{enumerate}
    \item{[1]\,\texttt{mergeid} --} ID of the merging system;
    \item{[2]\,\texttt{ncomp} --} number of components in this merging system;
    \item{[3--8]\,\texttt{id\_n} --} identification numbers of groups (\texttt{groupid}) that belong to this merging system.
    
\end{enumerate}

\end{document}